-----------------------------------------------------------------------------

\documentstyle[12pt]{article}

\begin{document}

\topmargin 0pt
\oddsidemargin 7mm
\headheight 0pt
\topskip 0mm
\addtolength{\baselineskip}{0.40\baselineskip}

\hfill SOGANG-HEP 191/94

\hfill September 15, 1994

\vspace{1cm}

\begin{center}
{\large \bf Batalin-Tyutin Quantization of the
            Self-Dual Massive Theory in Three Dimensions}
\end{center}

\vspace{1cm}

\begin{center}
Yong-Wan Kim, Young-Jai Park, Kee Yong Kim and Yongduk Kim \\
{\it Department of Physics and Basic Science Research Institute \\
Sogang University, C.P.O. Box 1142, Seoul 100-611, Korea}
\end{center}

\vspace{2cm}

\begin{center}
{\bf ABSTRACT}
\end{center}

We quantize the self-dual massive theory by using the Batalin-Tyutin
Hamiltonian method, which systematically embeds second class constraint
system into first class one in the extended phase space by introducing
the new fields.
Through this analysis we obtain simultaneously the
St\"uckelberg scalar term related to the explicit gauge-breaking effect and
the new type of Wess-Zumino action related to the Chern-Simons term.

\vspace{3cm}

\newpage

\begin{center}
\section{\bf Introduction}
\end{center}

The Dirac method has been widely used in the Hamiltonian formalism [1]
to quantize the second class constraint system.
However, since the resulting Dirac brackets are generally field-dependent
and nonlocal, and have a serious ordering problem
between field operators, these are under unfavorable circumstances
in finding canonically conjugate pairs.
On the other hand,
the quantizations of the first class  constraint systems [2,3] have been well
appreciated in a gauge invariant manner preserving
Becci-Rouet-Stora-Tyutin (BRST) symmetry [4,5].
If the second class constraint
system can be converted into first class one in an extended phase space,
we do not
need to define Dirac brackets and then the remaining quantization
program follows the method of Ref. [2-5].
This procedure has been extensively studied by Batalin, Fradkin, and Tyutin
[6,7] in the canonical formalism, and applied to various models [8-10]
obtaining the Wess-Zumino (WZ) action [11,12].

Recently, Banerjee [13] has applied the
Batalin-Tyutin (BT) Hamiltonian method [7]
to the second class constraint system of
the abelian Chern-Simons (CS) field theory [14-16], which yields first class
constraint algebra in an extended phase space
by introducing the new fields.
As a result, he has obtained the new type of an abelian WZ action,
which cannot be obtained in the usual path-integral framework.
Very recently, we have quantized the nonabelian case, which yields the
weakly involutive first class system originating from the second class one,
by generalizing this BT formalism [17].
As shown in these works,
the nature of second class constraint algebra originates from the
symplectic structure of CS term, not due to the local gauge symmetry
breaking.
Banerjee, and Ghosh [18] have also considered a massive Maxwell theory,
which has the explicit gauge-breaking term, in the BT approach.
As a result, the extra field in this approach has identified
with the St\"uckelberg scalar.
There are some other interesting examples in this approach [19].

In the present paper, we shall apply the BT Hamiltonian method [7]
to the self-dual massive theory [20]
revealing both the St\"uckelberg effect [21] and CS effect [13,17] by using
the BT quantization [7,18].
In section 2, since the BT formalism has developed quite recently,
we first recapitulate this formalism by explicitly analyzing
the pure CS theory.
In section 3, we apply the results discussed
in section 2 to the well-known self-dual massive theory
including the CS term in three dimensions,
which is gauge non-invariant.
By identifying the new fields $\rho$ and $\lambda$ with the
St\"uckelberg scalar and WZ scalar, respectively,
we obtain
simultaneously the St\"uckelberg scalar term related
to the explicit gauge-breaking mass term
and the new type of WZ action, which also
includes the St\"uckelberg scalar in order to maintain the gauge invariance
related to the CS term.

\vspace{1cm}
\begin{center}
\section {\bf The BT Formalism - The Pure CS Model}
\end{center}

Now, we first recapitulate the BT formalism by analyzing
the pure abelian CS model
\begin{equation}
 S = \int d^3 x [- \frac{m}{2} \epsilon_{\mu\nu\rho}
         B^\mu \partial^\nu B^\rho].
\end{equation}
Since this action is invariant up to the total divergence
under the gauge transformation
$\delta B^{\mu} = \partial^{\mu} \Lambda$,
this action has a different origin of the second class constraint
from the well-known massive Maxwell theory [18],
which is due to the explicit gauge symmetry breaking term in the action.
The origin of the second class constraints
is due to the symplectic structure of the CS model.

Following the usual Dirac's standard procedure [1],
we find that there are three primary constraints
\begin{eqnarray}
\Omega_0 &\equiv& \pi_0 \approx 0, \nonumber \\
\Omega_i &\equiv& \pi_i + \frac{1}{2} m \epsilon_{ij} B^j \approx 0
                 ~~( i = 1, 2 ),
\end{eqnarray}
and one secondary constraint,
\begin{equation}
\omega_3 \equiv - m \epsilon_{ij} \partial^i B^j \approx 0,
\end{equation}
obtained by conserving $\Omega_0$ with the total Hamiltonian,
\begin{equation}
H_T = H_c + \int d^2x [ u^0 \Omega_0 + u^i \Omega_i ],
\end{equation}
where $H_c$ is the canonical Hamiltonian,
\begin{equation}
H_c = \int d^2x ~[  m \epsilon_{ij} B^0 \partial^i B^j ],
\end{equation}
and we denote $x=(t,\vec{x})$ and two-space vector
$\vec{x}=(x^1,x^2)$ and $\epsilon_{12}=\epsilon^{12}=1$
and Lagrange multipliers $u^0$, $u^i$.
No further constraints are generated via this iterative procedure.
We find that all rest constraints except $\Omega_0 = \pi_0 \approx 0$
are superficially second class constraints.
However, in order to extract out the true second class constraints,
it is essential to redefine $\omega_3$ by using $\Omega_1$ and
$\Omega_2$ as follows
\begin{eqnarray}
\Omega_3 &\equiv& \omega_3 + \partial^i \Omega_i \nonumber\\
         &=&     \partial^i \pi_i
               - \frac{1}{2} m \epsilon_{ij} \partial^i B^j.
\end{eqnarray}
Then, $\Omega_0$, $\Omega_3$ form the first class algebra, while
$\Omega_1$, $\Omega_2$ form the second class algebra as follows
\begin{eqnarray}
\Delta_{i j}(x,y) &\equiv&
                  \{ \Omega_{i}(x), \Omega_{j}(y) \} \nonumber\\
               &=&
          \left( \begin{array}{cc}
                  0         &    m        \\
                 -m         &    0           \\
           \end{array}
          \right)
  \delta^2(x-y) ;~~ i, j = 1, 2.
\end{eqnarray}

In order to convert this system into first class one,
the first objective is to transform $\Omega_i$ into the first class
by extending the phase space.
Following the BT approach [7],
we introduce new auxiliary fields $\Phi^i$ to convert the second
class constraint $\Omega_i$ into
the first class one in the extended phase space,
and assume that the Poisson algebra of the new fields is given by
\begin{equation}
   \{ \Phi^i(x), \Phi^j(y) \} = \omega^{ij}(x,y),
\end{equation}
where $\omega^{ij}$ is an antisymmetric matrix.
Then, the modified constraint in the extended phase space is given by
\begin{equation}
  \tilde{\Omega}_i(\pi_\mu, B^\mu, \Phi^i)
         =  \Omega_i + \sum_{n=1}^{\infty} \Omega_i^{(n)};
                       ~~~~~~\Omega_i^{(n)} \sim (\Phi^i)^n,
\end{equation}
satisfying the boundary condition,
$\tilde{\Omega}_i(\pi_\mu, B^\mu, \Phi^i = 0) = \Omega_i$.
The first order correction term in the infinite series [7] is given by
\begin{equation}
  \Omega_i^{(1)}(x) = \int d^2 y X_{ij}(x,y)\Phi^j(y),
\end{equation}
and the first class  constraint algebra of $\tilde{\Omega}_i$ requires the
condition as follows
\begin{equation}
   \triangle_{ij}(x,y) +
   \int d^2 w~ d^2 z~
        X_{ik}(x,w) \omega^{kl}(w,z) X_{lj}(z,y)
         = 0.
\end{equation}
As was emphasized in Ref. [13,17], there is a natural arbitrariness
in choosing $\omega^{ij}$ and $X_{ij}$ from Eq.(8) and Eq.(10),
which corresponds to the canonical transformation
in the extended phase space [6,7].
We take the simple solutions as
\begin{eqnarray}
\omega^{i j} (x,y) &=&
    \left( \begin{array}{cc}
                 0         &    -1           \\
                  1         &    0           \\
           \end{array}
   \right)
  \delta^2(x-y),      \nonumber \\
X_{i j} (x,y) &=&
      \left( \begin{array}{cc}
             \sqrt{m}         &    0          \\
             0         &    \sqrt{m}           \\
           \end{array}
   \right)
  \delta^2(x-y),
\end{eqnarray}
which are compatible with Eq.(11) as it should be.
Using Eqs.(9), (10) and (12), the new set of constraints is found to be
\begin{equation}
\tilde{\Omega}_i = \pi_i + \frac{1}{2} m \epsilon_{ij} B^j + \sqrt{m} \Phi^i,
{}~~~~~~~(i=1,2),
\end{equation}
which are strongly involutive,
\begin{equation}
\{ \tilde{\Omega}_{\alpha}, \tilde{\Omega}_{\beta} \} = 0~~~~~~
 (\alpha, \beta = 0, 1, 2, 3)
\end{equation}
$\mbox{with}~~\tilde{\Omega}_0 \equiv
\Omega_0~~\mbox{and}~~\tilde{\Omega}_3
\equiv \Omega_3$.
As a result, we have all first class constraints
in the extended phase space
by applying the BT formalism systematically.
Observe further that only $\Omega_i^{(1)}$ contributes in the series
(9) defining the first class constraint.
All higher order terms given by Eq.(9) vanish
as a consequence of the choice Eq.(12).

Next, we derive the corresponding involutive Hamiltonian
in the extended phase space.
It is given by the infinite series [7],
\begin{equation}
 \tilde{H} = H_c + \sum_{n=1}^{\infty} H^{(n)}; ~~~~~H^{(n)} \sim (\Phi^i)^n,
\end{equation}
satisfying the initial condition,
$\tilde{H}(\pi_\mu, B^\mu, \Phi^i = 0) = H_c$.
The general solution [7] for the involution of $\tilde{H}$ is given by
\begin{equation}
  H^{(n)} = -\frac{1}{n} \int d^2 x d^2 y d^2 z~
              \Phi^i(x) \omega_{ij}(x,y) X^{jk}(y,z) G_k^{(n-1)}(z),
              ~~~(n \geq 1),
\end{equation}
where the generating functions $G_k^{(n)}$ are given by
\begin{eqnarray}
  G_i^{(0)} &=& \{ \Omega_i^{(0)}, H_c \},  \nonumber  \\
  G_i^{(n)} &=& \{ \Omega_i^{(0)}, H^{(n)} \}_{\cal O}
                    + \{ \Omega_i^{(1)}, H^{(n-1)} \}_{\cal O}
                                       ~~~ (n \geq 1),
\end{eqnarray}
where the symbol ${\cal O}$ in Eq.(17) represents
that the Poisson brackets are calculated among the original variables, i.e.,
${\cal O}=(\pi_\mu, B^\mu)$.
Here, $\omega_{ij}$ and $X^{ij}$ are the inverse matrices of $\omega^{ij}$
and $X_{ij}$ respectively.
Explicit calculations yield,
\begin{equation}
G_i^{(0)} =  - m \epsilon_{ij} \partial^j B^0,
\end{equation}
which is substituted in Eq.(16) to obtain $H^{(1)}$,
\begin{equation}
H^{(1)} = \int d^2x
                  [- \sqrt{m} (\partial_i \Phi^i) B^0].
\end{equation}
In this case, there are no further iterative Hamiltonians. Thus,
the total corresponding canonical Hamiltonian is
\begin{equation}
\tilde H = H_c + H^{(1)},
\end{equation}
which is involutive with the first class constraints,
\begin{eqnarray}
\{\tilde{\Omega}_i, \tilde H\} &=& 0,~~~~~~~ ( i = 1, 2, 3), \nonumber \\
\{\tilde{\Omega}_0, \tilde H\} &=& \tilde{\Omega}_3.
\end{eqnarray}
This completes the operatorial conversion of the original
second class system with Hamiltonian $H_c$ and constraints
$\Omega_i$ into first class with the involutive Hamiltonian $\tilde H$
and constraints $\tilde{\Omega}_i$.
Note that this Hamiltonian naturally generates the first class
Gauss' law constraint $\tilde{\Omega}_3$ from the time evolution
of $\tilde{\Omega}_0$.

Let us identify the new variables
$\Phi^i$ as a canonically conjugate pair
($\lambda$, $\pi_\lambda$) in the Hamiltonian formalism,
\begin{eqnarray}
\Phi^1 &\to& \frac{1}{\sqrt{m}} \pi_{\lambda}, \nonumber\\
\Phi^2 &\to& \sqrt{m} \lambda
\end{eqnarray}
satisfying Eqs.(8) and (12).
Then, the starting phase space partition function is given
by the Faddeev formula [3,22] as follows
\begin{equation}
Z =  \int  {\cal D} B^\mu {\cal D} \pi_\mu
           {\cal D} \lambda {\cal D} \pi_\lambda
          \prod_{\alpha,\beta = 0}^{3} \delta(\tilde{\Omega}_\alpha)
                               \delta(\Gamma_\beta)
          det \mid \{\tilde{\Omega}_\alpha,\Gamma_\beta\} \mid
          e^{iS},
\end{equation}
where
\begin{equation}
S  =  \int d^3x \left( \pi_\mu {\dot B}^\mu + \pi_\lambda {\dot \lambda}
                 - \tilde{\cal H}
           \right),
\end{equation}
with Hamiltonian density $\tilde {\cal H}$ corresponding to Hamiltonian
$\tilde H$, which is now expressed in terms
of $(\lambda, \pi_\lambda)$ instead of $\Phi^i$.
The gauge fixing conditions $\Gamma_i$ are chosen
so that the determinant occurring in
the functional measure is nonvanishing.
Moreover, $\Gamma_i$ may be assumed to be independent of the momenta
so that these are considered as Faddeev-Popov type gauge conditions [23].

We now perform the momentum integrations to obtain the
configuration space partition function.
The  $\pi_0$, $\pi_1$, and $\pi_2$ integrations are
trivially performed by exploiting the delta function
$~\delta(\tilde{\Omega}_0)~ =~ \delta(\pi_0),~~~
\delta(\tilde{\Omega}_1)~ =~ \delta(\pi_1+\frac{m}{2}B^2+\pi_\lambda),
{}~~~ {\mbox {and}}
{}~~~\delta(\tilde{\Omega}_2)~ = \\ ~\delta(\pi_2-\frac{m}{2}B^1+m\lambda),$
respectively.
After exponentiating the remaining delta function
$\delta(\tilde{\Omega}_3)
= \delta(-m\epsilon_{ij}\partial^iB^j
+\partial_1\pi_\lambda+m\partial_2\lambda)$
with Fourier variable $\xi$
and transforming $B^0 \to B^0 + \xi$,
we finally obtain the action as follows
\begin{equation}
S = \int d^3x
       [  - \frac{1}{2} m \epsilon_{\mu\nu\rho} B^\mu \partial^\nu B^\rho
               + m \lambda  F_{02} ],
\end{equation}
where $F_{02}=\partial_0 B_2 - \partial_2 B_0$,
and the corresponding Liouville measure
just comprises the configuration space variables as follows
\begin{equation}
[{\cal D} \mu] = {\cal D} B^\mu
                 {\cal D} \lambda
                 {\cal D} \xi
                 \delta(F_{01}+\dot{\lambda})
                 \prod^3_{\beta = 0}
               \{ \delta(\Gamma_{\beta}[B^0 + \xi, B^i, \lambda]) \}
                det \mid \{\tilde{\Omega}_{\alpha}, \Gamma_{\beta}\} \mid,
\end{equation}
where $\delta(F_{01}+\dot{\lambda})$ is
expressed by $\int {\cal D} \pi_\lambda
e^{-i\int d^3x ~(F_{01} + \dot{\lambda}) \pi_\lambda}$.
This action is invariant up to the total divergence
under the gauge transformations as
$\delta B_\mu = \partial_\mu \Lambda$ and $\delta \lambda = 0$.
It is easily checked for consistency that
starting from the Lagrangian (25) with a factor
$\delta(F_{01}+\dot{\lambda})$ in the measure part,
one can exactly reproduce the set of all first class constraints
$\tilde{\Omega}_\alpha$ and the involutive Hamiltonian (20).
Note that we will show in the next section that the above
$\delta$-function, which is remaining in the measure part for
the case of the pure CS theory,
will be disappeared for the
case of the non-pure CS theories [13,17]
like the self-dual massive theory [20].

\vspace{1cm}

\begin{center}
\section {\bf The Self-Dual Massive Model}
\end{center}

We consider the Abelian self-dual massive model [20]
\begin{equation}
 S_{SD}  =  \int d^3 x~ [ \frac{1}{2} m^2 B^\mu B_\mu
             - \frac{1}{2} m \epsilon_{\mu\nu\rho}
             B^{\mu} \partial^{\nu} B^{\rho}],
\end{equation}
by using the useful results discussed in the previous section.
Note that this action has an explicit mass term,
which breaks the gauge symmetry as the case of the Proca model,
and also the CS term, which has a different origin of the second class
constraint system.
Consequently, this action represents a second class constraint system,
which can be easily confirmed by the standard constraint analysis.
There are three primary constraints,
\begin{eqnarray}
\Omega_0 &\equiv& \pi_0 \approx 0, \nonumber \\
\Omega_i &\equiv& \pi_i + \frac{1}{2} m \epsilon_{ij} B^j \approx 0,
                 ~~~~~( i = 1, 2 ),
\end{eqnarray}
and one secondary constraint,
\begin{equation}
\omega_3 \equiv m^2 B^0 - m \epsilon_{ij} \partial^i B^j \approx 0,
\end{equation}
which is obtained by conserving $\Omega_0$ with the total Hamiltonian,
\begin{equation}
H_T = H_c + \int d^2x [ u^0 \Omega_0 + u^i \Omega_i ],
\end{equation}
where $H_c$ is the canonical Hamiltonian,
\begin{equation}
H_c = \int d^2x \left[  \frac{1}{2} m^2 \{ (B^i)^2 - (B^0)^2 \}
                      + m \epsilon_{ij} B^0 \partial^i B^j \right],
\end{equation}
and  $u^0$ and $u^i$ are Lagrange multipliers.
No further constraints are generated via this iterative procedure.
We find that all constraints are fully second class constraints.
It is, however, essential to redefine $\omega_3$ by using $\Omega_1$ and
$\Omega_2$ as follows
\begin{eqnarray}
\Omega_3 &\equiv& \omega_3 + \partial^i \Omega_i \nonumber\\
         &=&     \partial^i \pi_i
               - \frac{1}{2} m \epsilon_{ij} \partial^i B^j
               + m^2 B^0,
\end{eqnarray}
although the redefined constraints are still completely second class
in contrast to the case of the pure CS theory.
Otherwise, one will have a complicated constraint algebra including the
derivative terms, which is difficult
to handle.
Then, the constraint algebra is given by
\begin{eqnarray}
\Delta_{\alpha\beta}(x,y) &\equiv&
                  \{ \Omega_{\alpha}(x), \Omega_{\beta}(y) \} \nonumber\\
               &=&
          \left( \begin{array}{cccc}
         0 &          0         &    0   &    - m^2    \\
         0 &          0         &    m   &     0        \\
         0 &         -m         &    0   &     0        \\
         m^2 &        0         &    0    &     0
           \end{array}
          \right)
  \delta^2(x-y) ;  \alpha, \beta = 0, 1, 2, 3,
\end{eqnarray}
which reveals the simple second class nature
of the constraints $\Omega_\alpha (x)$.

In order to convert this system into the first class one,
the first objective is
to transform $\Omega_\alpha$ into the first class
by extending the phase space.
Following the BT approach [7], we introduce the matrix (8) as follows
\begin{equation}
\omega^{\alpha\beta} (x,y) =
    \left( \begin{array}{cccc}
         0 &          0         &    0   &    1    \\
         0 &          0         &    -1   &     0        \\
         0 &          1         &    0   &     0        \\
        -1 &        0         &    0    &     0
           \end{array}
   \right)
  \delta^2(x-y).
\end{equation}
Then the other matrix $X_{\alpha\beta}$ in Eq.(10) is obtained by
solving Eq.(11) with $\Delta_{\alpha\beta}$ given by Eq.(33),
\begin{equation}
X_{\alpha\beta} (x,y) =
      \left( \begin{array}{cccc}
         m &          0         &    0   &    0    \\
         0 &          \sqrt{m}         &    0   &     0        \\
         0 &          0         &    \sqrt{m}   &     0        \\
         0 &          0         &    0    &     m
           \end{array}
   \right)
  \delta^2(x-y).
\end{equation}
There is an arbitrariness in choosing $\omega^{\alpha\beta}$,
which would naturally be manifested in Eq.(34). This just corresponds
to the canonical transformations in the extended phase space.
However, as has also been evidenced in other calculations [13,17],
this choice of Eqs.(34) and (35) brings about
remarkable algebraic simplifications.

Using Eqs.(9), (10) and (35), the new set of constraints is found to be
\begin{eqnarray}
\tilde{\Omega}_0 &=& \pi_0 + m \Phi^0,  \nonumber\\
\tilde{\Omega}_i &=& \pi_i + \frac{1}{2} m \epsilon_{i j} B^j
                     + \sqrt{m} \Phi^i, ~~~~~~~~(i=1,2), \nonumber\\
\tilde{\Omega}_3 &=& \partial^i \pi_i
                    - \frac{1}{2} m \epsilon_{ij} \partial^i B^j
                    + m^2 B^0
                    + m \Phi^3,
\end{eqnarray}
which are strongly involutive,
\begin{equation}
\{ \tilde{\Omega}_{\alpha}, \tilde{\Omega}_{\beta} \} = 0.
\end{equation}
Recall the $\Phi^{\mu}$ are the new variables satisfying the algebra
(8) with $\omega^{\alpha\beta}$  given  by Eq.(34).

The next step is to obtain the involutive Hamiltonian.
The generating functional $G_\alpha^{(n)}$
is obtained from Eq.(17).
It is noteworthy that there are only two terms $\Omega_\alpha$ and
$\tilde{\Omega}_\alpha^{(1)}$ in the expansion (36)
due to the intuitive choice (34) and (35).
Explicit calculations yield,
\begin{eqnarray}
G_0^{(0)} &=& m^2 B^0 - m \epsilon_{ij} \partial^i B^j, \nonumber \\
G_i^{(0)} &=& - m^2 B^i - m \epsilon_{ij} \partial^j B^0, \nonumber \\
G_3^{(0)} &=& m^2 \partial_i B^i,
\end{eqnarray}
which are substituted in Eq.(16) to obtain $H^{(1)}$,
\begin{equation}
H^{(1)} = \int d^2x  \left[
                  m \Phi^0 \partial_i B^i
                  + m \sqrt{m} \epsilon_{ij} \Phi^i B^j
                  + \sqrt{m} \Phi^i \partial_i B^0
                  - \Phi^3 ( m B^0 - \epsilon_{ij} \partial^i B^j )
               \right].
\end{equation}
This is inserted back in Eq.(17) to deduce $G_\alpha^{(1)}$ as follows
\begin{eqnarray}
G_0^{(1)} &=& {\sqrt m} \partial_i \Phi^i  + m \Phi^3, \nonumber \\
G_i^{(1)} &=& m \partial_i \Phi^0 + m {\sqrt m} \epsilon_{ij} \Phi^j
                - \epsilon_{ij} \partial^j \Phi^3, \nonumber\\
G_3^{(1)} &=& m \partial_i \partial^i \Phi^0 + m {\sqrt m} \epsilon_{ij}
                \partial^i \Phi^j,
\end{eqnarray}
which then yield $H^{(2)}$ from Eq.(16),
\begin{equation}
H^{(2)} = \int d^2x  \left[
                 - \frac{1}{2} \partial_i \Phi^0 \partial^i \Phi^0
                 + \sqrt{m} \Phi^0 \epsilon_{ij} \partial^i \Phi^j
                 + \frac{1}{2} m \Phi^i \Phi^i
    - ( \frac{1}{\sqrt{m}} \partial_i \Phi^i + \frac{1}{2} \Phi^3 ) \Phi^3
       \right].
\end{equation}
Since $G_\alpha^{(n)} = 0 ~~(n \geq 2)$, the final expression for the
desired involutive Hamiltonian after the $n=2$ finite truncations
is given by
\begin{equation}
\tilde H = H_c + H^{(1)} + H^{(2)},
\end{equation}
which, by construction, is involutive,
\begin{equation}
\{\tilde{\Omega}_\alpha, \tilde H\} = 0.
\end{equation}
This completes the operatorial conversion of the original
second class system with Hamiltonian $H_c$ and constraints
$\Omega_\alpha$ into first class
with Hamiltonian $\tilde H$ and constraints $\tilde{\Omega}_\alpha$.

Before performing the momentum integrations to obtain the
partition function in the configuration space,
it seems appropriate to comment on the involutive Hamiltonian.
If we directly use the above Hamiltonian,
we will finally obtain the non-local action corresponding to this Hamiltonian
due to the existence of
$\frac{1}{\sqrt{m}}\Phi^3\partial_1\Phi^1$--term in the action
when we carry out the functional integration over $\Phi^1$ or $\Phi^3$ later.
Furthermore, if we use the above Hamiltonian,
we can not also naturally generate
the first class Gauss' law constraint $\tilde{\Omega}_3$ from
the time evolution of the primary constraint $\tilde{\Omega}_0$,
which is the first class.
Therefore, in order to avoid these problems,
we use the equivalent first class Hamiltonian
without any loss of generality,
which differs from the involutive Hamiltonian (42)
by adding a term proportional to the first class constraint
$\tilde{\Omega}_3$ as follows
\begin{equation}
\tilde{H}^{'} = \tilde{H} + \frac{1}{m} \Phi^3 \tilde{\Omega}_3.
\end{equation}
Then, this Hamiltonian $\tilde {H}^{'}$ consistently generates
the Gauss' law constraint
such that $\{ \tilde{\Omega}_0, \tilde{H}^{'} \} = \tilde{\Omega}_3$.
Note that non-locality may also be avoided by changing the order
of performing the momentum integrals.
But, in this case, one can not directly reproduce the original
theory by fixing the unitary gauge as well as the Gauss' law constraint.
Furthermore, when we act this modified Hamiltonian on physical states,
the difference with $\tilde{H}$ is trivial because such states are
annihilated by the first class constraints.
Similarly, the equations of motion for observable ({\it i.e.}
gauge invariant variables) will also be unaffected by this difference
since $\tilde{\Omega}_3$ can be regarded as
the generator of the gauge transformations.

We now unravel the correspondence of the Hamiltonian approach including
both the St\"uckelberg effect and the CS effect.
The first step is to identify the new variables
$\Phi^\mu$ as canonically conjugate pairs in the Hamiltonian formalism,
\begin{eqnarray}
\Phi^0 &\to& m \rho,   \nonumber\\
\Phi^1 &\to& \frac{1}{\sqrt{m}} \pi_{\lambda}, \nonumber\\
\Phi^2 &\to& \sqrt{m} \lambda, \nonumber\\
\Phi^3 &\to& \frac{1}{m} \pi_{\rho},
\end{eqnarray}
satisfying Eqs.(8), (34) and (35).
The starting phase space partition function is then given
by the Faddeev formula,
\begin{equation}
Z =  \int  {\cal D} B^\mu {\cal D} \pi_\mu
           {\cal D} \lambda {\cal D} \pi_\lambda
           {\cal D} \rho {\cal D} \pi_\rho
          \prod_{\alpha,\beta = 0}^{3} \delta(\tilde{\Omega}_\alpha)
                               \delta(\Gamma_\beta)
          det \mid \{\tilde{\Omega}_\alpha,\Gamma_\beta\} \mid
          e^{iS'},
\end{equation}
where
\begin{equation}
S'  =  \int d^3x \left( \pi_\mu {\dot B}^\mu + \pi_\lambda {\dot \lambda}
                 + \pi_\rho {\dot \rho}
                 - \tilde{\cal H'}
           \right)
\end{equation}
with the Hamiltonian density $\tilde{\cal H}'$ corresponding
to $\tilde H'$,
which is now expressed in
terms of  $(\rho, \pi_{\rho}, \lambda, \pi_\lambda)$
instead of $\Phi^\mu$.
As in the previous section, the gauge fixing
conditions $\Gamma_\alpha$ may be assumed to be independent
of the momenta so that these are considered  as  Faddeev-Popov  type
gauge conditions.

Next, we perform the momentum integrations to obtain the
configuration space partition function.
The  $\pi_0$, $\pi_1$, and $\pi_2$ integrations are
trivially performed by exploiting the delta functions
$~~\delta(\tilde{\Omega}_0)~ =~ \delta(\pi_0 + m^2 \rho)$,
$~~\delta(\tilde{\Omega}_1)~ =~ \delta(\pi_1+\frac{m}{2}B^2+\pi_\lambda)$,
and
$~~\delta(\tilde{\Omega}_2)~ =~ \delta(\pi_2-\frac{m}{2}B^1+m\lambda)$,
respectively.
After exponentiating the remaining delta function
$\delta(\tilde{\Omega}_3) = \delta(-m\epsilon_{ij}\partial^iB^j
+\partial_1\pi_\lambda+m\partial_2\lambda+m^2B^0+\pi_\rho)$
with Fourier variable $\xi$ as
$\delta(\tilde{\Omega}_3)=\int{\cal D}\xi e^{-i\int d^3x
\xi\tilde{\Omega}_3}$
and transforming $B^0 \to B^0 + \xi$,
we obtain the action as follows
\begin{eqnarray}
S &=& \int d^3x~ \{ \frac{1}{2} m^2 B^\mu B_\mu
 - \frac{1}{2} m \epsilon_{\mu\nu\rho} B^\mu \partial^\nu B^\rho \nonumber\\
     &+& \rho [ -m^2 (\dot{B^0} + \dot{\xi}) - m^2 \partial_i B^i
           - \frac{1}{2} m^2 \partial_i \partial^i \rho
           + m^2 \partial_1 \lambda - m \partial_2 \pi_\lambda ] \nonumber\\
     &+& \pi_\rho [ \dot{\rho} - \frac{1}{2m^2} \pi_\rho - \xi ]
           + \lambda [ - m \dot{B}^2 + m^2 B^1 - m \partial_2 B^0
           - \frac{1}{2} m^2 \lambda ] \nonumber\\
     &+&  \pi_\lambda [ \dot{\lambda} - \dot{B}^1
                      - m B^2 + \partial^1 B^0
          - \frac{1}{2} \pi_\lambda ] - \frac{1}{2} m^2 \xi^2  \},
\end{eqnarray}
and the corresponding measure is given by
\begin{equation}
[{\cal D} \mu] = {\cal D} B^\mu
                 {\cal D} \lambda
                 {\cal D} \pi_\lambda
                 {\cal D} \rho
                 {\cal D} \pi_\rho
                 {\cal D} \xi
                 \prod^3_{\beta = 0}
               \{ \delta(\Gamma_{\beta}[B^0 + \xi, B^i, \lambda, \rho]) \}
                det \mid \{\tilde{\Omega}_{\alpha}, \Gamma_{\beta}\} \mid,
\end{equation}
where
$B^0 \to B^0 + \xi$ transformation is naturally understood
in the gauge fixing condition $\Gamma_{\beta}$.

Note that the original theory is easily reproduced in one line,
if we choose the unitary gauge
\begin{equation}
\Gamma_0 = \rho,~~\Gamma_1 = \pi_\lambda,~~\Gamma_2 = \lambda,
               ~~\Gamma_3 = \pi_\rho,
\end{equation}
and integrate over $\xi$.
Then, one can easily realize that
the new fields $\Phi^\mu$ are nothing but the gauge degrees of
freedom, which can be removed by utilizing the gauge symmetry.

Now, we perform the Gaussian integration over $\pi_\rho$.
Then all $\xi$ terms in the action are canceled out,
and integrating over $\pi_\lambda$, the resultant action
is finally obtained as follows
\begin{eqnarray}
S &=& S_{St\ddot{u}ckelberg} + S_{WZ};  \nonumber \\
S_{St\ddot{u}ckelberg} &=& \int d^3x
        \{  \frac{1}{2} m^2 (B_\mu B^\mu
            + \partial_\mu \rho \partial^\mu \rho
            - 2 \rho \partial_\mu B^\mu )
             - \frac{1}{2} m \epsilon_{\mu\nu\rho}
             B^{\mu} \partial^{\nu} B^{\rho}
             \},
             \nonumber \\
          &=& \int d^3x
        \{  \frac{1}{2} m^2 (B_\mu + \partial_\mu \rho )^2
             - m^2 \partial_\mu (\rho B^\mu )
             - \frac{1}{2} m \epsilon_{\mu\nu\rho}
             B^{\mu} \partial^{\nu} B^{\rho}
             \},
             \nonumber \\
S_{WZ}  &=& \int d^3x
         \{ \frac{1}{2} [ {\dot \lambda} + F_{01}
                         + m (B_2 + \partial_2 \rho) ]^2 \nonumber \\
       &&~~~~~~~~~+~ m \lambda [ F_{02} - m (B_1 + \partial_1 \rho)
                         - \frac{1}{2} m \lambda ]    \},
\end{eqnarray}
where $F_{01}=\partial_0 B_1 - \partial_1 B_0$,
$F_{02}=\partial_0 B_2 - \partial_2 B_0$,
and the corresponding Liouville measure just comprises
the configuration space variables as follows
\begin{equation}
[{\cal D} \mu] = {\cal D} B^\mu
                 {\cal D} \lambda
                 {\cal D} \rho
                 {\cal D} \xi
                 \prod^3_{\beta = 0}
               \{ \delta(\Gamma_{\beta}[B^0 + \xi, B^i, \lambda, \rho]) \}
                det \mid \{\tilde{\Omega}_{\alpha}, \Gamma_{\beta}\} \mid.
\end{equation}
This action $S$ is invariant up to the total divergence
under the gauge transformations as
$\delta B_\mu = \partial_\mu \Lambda$, $\delta \rho = - \Lambda$, and
$\delta \lambda = 0$.
Starting from the Lagrangian (51) with the boundary term,
we can easily reproduce the same set of all first class constraints
$\tilde{\Omega}_\alpha$, and the Hamiltonian such that
\begin{eqnarray}
H = H_c &+& \int d^2x~~ [ \pi_\lambda \partial_1 B_0 - m \pi_\lambda
      + m^2 \rho \partial_i B^i + m \lambda \partial_2 B_0
      + m^2 \lambda B_1         \nonumber \\
  &+& \frac{1}{2} \pi_\lambda^2 - m \pi_\lambda \partial_2 \rho
      - \frac{1}{2} m^2 \partial_i \rho \partial^i \rho
      + m^2 \lambda \partial_1 \rho + \frac{1}{2} m^2 \lambda^2
      + \frac{1}{2} \pi_\rho^2].
\end{eqnarray}
Then, if we add a term proportional to the constraint
$\tilde{\Omega}_3$, $i.e., - \frac{1}{m^2} \pi_\rho \tilde{\Omega}_3$,
which is trivial when acting on the physical Hilbert space,
to the above Hamiltonian (53),
we can obtain the original involutive Hamiltonian (42),
which is canonically equivalent to Eq.(53).
Furthermore, this difference is also trivial in the construction
of the functional integral because the constraint $\tilde{\Omega}_3$
is strongly implemented by the delta function
$\delta(\tilde{\Omega}_3)$ in Eq.(46).
Therefore, we have shown that the constraints and Hamiltonian
following from the Lagrangian (51) are effectively equivalent to
the original Hamiltonian embedding.
As results, through BT quantization procedure,
we have found that
the St\"uckelberg scalar $\rho$ is naturally introduced in the mass term,
and this $\rho$ as well as the WZ scalar $\lambda$ is also included
in the new type of WZ action.

Note that the gauge invariance of $S_{\rm WZ}$ should be maintained
because the second-class constraint structure related to CS term only
comes from the symplectic structure.
We also note that the Wess-Zumino action in Eq.(51) is gauge invariant
in spite of the lack of the manifest Lorentz invariance.
On the other hand, since the unitary gauge (50) recovers
the manifestly Lorentz invariant original action,
the actual invariance is maintained from the fact that the final result
for the partition function $Z$ is independent of the gauge fixing conditions.
The local gauge symmetry of
the Wess-Zumino action naturally also survives in the configuration space.
This means that the origin of $S_{WZ}$ is irrelevant to the conventional
gauge-variant Wess-Zumino like action [8,10,11],
which cancels the local gauge anomaly of the second-class system.
Interestingly the choice of $\rho=0$ and $\lambda$=0
does not recover the original theory in the Faddeev-Popov type gauges.
Finally, it seems appropriate to comment on the action (51).
If we ignore the boundary term in the Lagrangian (51),
we cannot directly obtain the involutive first class Hamiltonian
as the case of the Proca theory explained in Ref. [18]
because the boundary term plays the important role in this procedure.

In summary, we have recapitulated the
Batalin-Tyutin method,
which converts second class system into first class one,
by analyzing the pure CS theory, which has the
different origin of the second class structure. Then, we have
applied this method to the Abelian self-dual massive theory
including the CS term.
As results, we have shown that if we ignore the boundary term
in action (51),
the direct connection with the usual Lagrangian embedding of
St\"uckelberg can be made by explicitly evaluating the momentum
integrals in the extended phase space partition function using
Faddeev-Popov-like gauges
and identifying an extra field $\rho$ introduced
in our Hamiltonian formalism with the conventional St\"uckelberg scalar.
On the other hand, we have also obtained a new type of
Wess-Zumino action containing the WZ scalar $\lambda$, which also includes
the St\"uckelberg scalar $\rho$
in order to maintain the gauge invariance of the
$S_{WZ}$ related to the CS effect in the action (51).

\vspace{1cm}

\begin{center}
\section*{Acknowledgements}
\end{center}

We would like to thank W. T. Kim and S. -K. Kim for helpful discussions.
The present study was supported in part by
the Basic Science Research Institute Program,
Ministry of Education, 1994, Project No. 2414,
and the Korea Science and Engineering Foundation through
the Center for Theoretical Physics.

\end{document}